\documentclass{llncs}

\usepackage{breakurl} % Not needed if you use pdflatex only.
\usepackage{makeidx}
\usepackage{graphicx,latexsym,alltt}
\usepackage{amssymb,amsmath}
\usepackage{algorithm,algorithmic}
\usepackage{subfigure}
\usepackage{color}
\usepackage{url}
\usepackage{listings}
\usepackage{multirow} 
\usepackage{enumitem} %Para cambiar la tabulacion de los 

\usepackage{multicol}

\newcommand{\biblio}[1]{../../../../../../../Biblio/#1}
\definecolor{blank}{rgb}{0.55,0.55,0.55}

\long\def\comment#1{}

\renewcommand{\phi}{\varphi}

\def\defemb#1#2{\expandafter\def\csname #1\endcsname
                              {\relax\ifmmode #2\else\hbox{$#2$}\fi}}
\defemb{cA}{{\cal A}}
\defemb{cB}{{\cal B}}
\defemb{cC}{{\cal C}}
\defemb{cD}{{\cal D}}
\defemb{cE}{{\cal E}}
\defemb{cF}{{\cal F}}
\defemb{cG}{{\cal G}}
\defemb{cH}{{\cal H}}
\defemb{cI}{{\cal I}}
\defemb{cJ}{{\cal J}}
\defemb{cL}{{\cal L}}
\defemb{cM}{{\cal M}}
\defemb{cN}{{\cal N}}
\defemb{cO}{{\cal O}}
\defemb{cP}{{\cal P}}
\defemb{cR}{{\cal R}}
\defemb{cS}{{\cal S}}
\defemb{cT}{{\cal T}}
\defemb{cU}{{\cal U}}
\defemb{cV}{{\cal V}}
\defemb{cW}{{\cal W}}
\defemb{cX}{{\cal X}}
\defemb{cZ}{{\cal Z}}

\makeatletter
\newenvironment{prog}{\vspace{1.0ex}\par
\obeylines\@vobeyspaces\tt}{\vspace{1.0ex}\noindent
}
\makeatother
\newcommand{\startprog}{\begin{prog}}
\newcommand{\stopprog}{\end{prog}\noindent}
\catcode`\_=\active
\let_=\sb
\catcode`\_=12
\makeatother

\newif\ifpaperVersion
\paperVersiontrue
\ifpaperVersion
	\newcommand{\deleted}[1]{}
	
	\newcommand{\pending}[1]{}
	\newcommand{\ignored}[1]{}
	\newcommand{\done}[1]{}
	\newcommand{\doubt}[1]{}
	\newcommand{\josep}[1]{}
	\newcommand{\david}[1]{}
	\newcommand{\sergio}[1]{}
	\newcommand{\tama}[1]{}
\else
	\definecolor{ignoredColor}{rgb}{1.0,0.75,0.25}
	\definecolor{pendingColor}{rgb}{0.0,0.0,1.0}
	\definecolor{doneColor}{rgb}{0.0,0.5,0.0}
	\definecolor{doubtColor}{rgb}{0.6,0.6,0.4}
	\definecolor{josepColor}{rgb}{0.2,0.6,0.6}
	\definecolor{davidColor}{rgb}{0.6,0.2,0.6}
	\definecolor{sergioColor}{rgb}{1.0,0.5,0.0}
	\definecolor{tamaColor}{rgb}{0.54, 0.27, 0.07}

	\newcommand{\deleted}[1]{\textcolor{red}{\#Deleted: #1}}
	
	\newcommand{\pending}[1]{\textcolor{pendingColor}{\#Pending: #1}}
	\newcommand{\ignored}[1]{\textcolor{ignoredColor}{\#Ignored: \{\{ #1 \}\}}}
	\newcommand{\done}[1]{\textcolor{doneColor}{\#Done: \{\{ #1 \}\}}}
	\newcommand{\doubt}[1]{\textcolor{doubtColor}{\#Doubt: #1}}
	\newcommand{\josep}[1]{\textcolor{josepColor}{\#JJJ: #1}}
	\newcommand{\david}[1]{\textcolor{davidColor}{\#DDD: #1}}
	\newcommand{\sergio}[1]{\textcolor{sergioColor}{\#SSS: #1}}
	\newcommand{\tama}[1]{\textcolor{tamaColor}{\#TTT: #1}}
\fi

\usepackage{listings}
\usepackage[utf8]{inputenc}
\usepackage{courier}
\usepackage[T1]{fontenc}
\usepackage[scaled]{couriers}

\usepackage[utf8]{inputenc}
\usepackage[T1]{fontenc}
\usepackage[svgnames]{xcolor}
\usepackage[scaled=0.84]{beramono}
\usepackage{listings}

\makeatletter
\newcommand{\MyChange}[2]{%
  \extractcolorspec{.}\MyChange@CurrentColor
  \extractcolorspec{#2}\MyChange@TestColor
  \ifx\MyChange@CurrentColor\MyChange@TestColor
    {\bfseries\color{green!33!black} #1 }%
  \else
    {#1 }%
  \fi%
}
\makeatother

\lstdefinelanguage{Erlang}
{
classoffset=1,
keywords={},
  keywords = [2]{receive, if, case, try, module, export, spawn, end, fun, when, after, spec, of},
  keywordstyle=\textbf,
  keywordstyle=[2]\color{blue},% for example
classoffset=2,
  otherkeywords={% Operators
    ;, ",\,,\#, ., (, ), \{, \}, [,],|,||,+,-,*,/, <,>, >=, =<,->, =, ==, =:=, _
  },
  classoffset=0,
  sensitive=true,
   morecomment=[l]{\%},
     commentstyle=\color{red}\emph,
  stringstyle=\color{black},
  morestring=[b]',
}

\lstdefinelanguage{none}
{
keywords={},
keywords = [2]{},
otherkeywords={},
morekeywords={}
  ,classoffset=1
  ,sensitive=true
  ,stringstyle=\color{black}
  ,morestring=[b]',
  numbers=none,
  ,literate=
}

\begin{document}

\frontmatter
\pagestyle{headings}
\addtocmark{Hamiltonian Mechanics}

\pagenumbering{arabic}

% Titulo
\title
{
 Erlang Code Evolution Control\\(Use Cases)
\thanks
{
This work has been partially supported by MINECO/AEI/FEDER (EU)
under grant TIN2016-76843-C4-1-R and by the \emph{Generalitat
Valenciana} under grant PROMETEO-II/2015/013 (SmartLogic).
%Sergio P\'erez was partially supported by the Spanish \emph{Iniciativa
%de Empleo Juvenil} Programme (MINECO) and the European Social Fund in
%collaboration with the \emph{Sistema Nacional de Garant\'{\i}a Juvenil} under
%grant PEJ-2014-A-24709 (\emph{Promoci\'on de Empleo Joven e Implantaci\'on
%de la Garant\'{\i}a Juvenil 2014}, MINECO).
Salvador Tamarit was partially supported by the \emph{Conselleria de
Educaci\'on, Investigaci\'on, Cultura y Deporte de la Generalitat
Valenciana} under grant APOSTD/2016/036.
}
}
\titlerunning{Erlang Code Evolution Control\\(Uses Cases)}

% Autores
\author{David Insa \and Sergio P\'erez \and Josep Silva \and Salvador Tamarit}
\institute
{
%Departament de Sistemes Inform\`atics i Computaci\'o\\
Universitat Polit\`ecnica de Val\`encia\\
Cam\'i de Vera s/n, E-46022 Val\`encia, Spain\\
 \email{\{dinsa,serperu,jsilva,stamarit\}@dsic.upv.es}
%\and
%%Babel Research Group\\
%Universidad Polit\'ecnica de Madrid\\
%Campus de Montegancedo s/n, E-28660 Boadilla del Monte, Spain\\
% \email{salvador.tamarit@upm.es}
}

% Genera el titulo
\maketitle

% Abstract
%!TEX root = ../paper.tex

\begin{abstract}
The main goal of this work is to show how SecEr can be used in different scenarios. Concretely, we demonstrate how a user can run SecEr to obtain reports about the behaviour preservation between versions as well as how a user can use SecEr to find the source of a discrepancy. The use cases presented are three: two completely different versions of the same program, an improvement in the performance of a function  and a program where an error has been introduced. A complete description of the technique and the tool is available at \cite{insa2017LOPSTR}  and \cite{insa2017LOPSTR_LNCS}.
\end{abstract}

% Keywords

% Secciones
%\input{./Contenido.tex}
%!TEX root = ./paper.tex

%%%%%%%%%%%%%%%%%%%%%%%%%%%%%%%%%%%%%%%%%%%%%%%%%%%%%%%%%%%%%%%%%%
%\section{Use cases}
%\label{sec:app-use-cases}

\section{Use case 1: Happy numbers}\label{sec:uc_happy}

In order to show the potential of the tool, we provide an example to compare two versions of an Erlang program that computes happy numbers. Concretely, these two versions are based on completely different algorithms. Therefore, this use case shows that SecEr is able to deal with such scenarios where the correspondence between versions is not direct. Both of them are taken from the Rosetta Code repository:\footnote{Consulted in this concrete version: \url{http://rosettacode.org/mw/index.php?title=Happy_numbers&oldid=251560#Erlang}}
\begin{center}
\footnotesize
\url{http://rosettacode.org/wiki/Happy_numbers#Erlang}.
\end{center}

\begin{lstlisting}[float, language=none]
\end{lstlisting}

\setlength{\columnsep}{1cm}
\noindent\begin{minipage}{\linewidth}
\begin{multicols}{2}
\begin{lstlisting}[tabsize=2,basicstyle=\ttfamily\tiny, frame=single, caption={happy0.erl}, label=lst:happy0, numbers=left, stepnumber=1,escapechar=@]
-spec main(pos_integer(),pos_integer()) -> 
	[pos_integer()].
main(N, M) -> 
	happy_list(N, M, []).

happy_list(_, N, L) when length(L) =:= N -> 
	lists:reverse(L);
happy_list(X, N, L) -> 
	Happy = is_happy(X),@\label{lst:poi_happy0}@
 	if Happy -> 
		happy_list(X + 1, N, [X|L]);
	true -> 
		happy_list(X + 1, N, L) end.

is_happy(1) -> true;
is_happy(4) -> false;
is_happy(N) when N > 0 ->
	N_As_Digits = 
		[Y - 48 || 
		Y <- integer_to_list(N)],
	is_happy(
		lists:foldl(
			fun(X, Sum) -> 
				(X * X) + Sum 
			end, 
			0, 
			N_As_Digits));
is_happy(_) -> false.
\end{lstlisting}
\columnbreak
\begin{lstlisting}[tabsize=2,basicstyle=\ttfamily\tiny, frame=single, caption={happy1.erl}, label=lst:happy1, numbers=left, stepnumber=1,escapechar=@]
is_happy(X, XS) ->
 	if@\label{lst:poi_happy1_ex2_1}@
 		X == 1 -> true;
 		X < 1 -> false;@\label{lst:bug_1}@
 		true ->
 			case member(X, XS) of
 				true -> false;
 				false ->
 					is_happy(sum(map(fun(Z) -> Z*Z end, 
						[Y - 48 || Y <- integer_to_list(X)])),
					  [X|XS])
			end
	end.
happy(X, Top, XS) ->
	if@\label{lst:poi_happy1_ex2_2}@
		length(XS) == Top -> sort(XS);
		Happy = is_happy(X,[]),@\label{lst:poi_happy1}@
		true ->
			case  Happy of
				true -> happy(X + 1, Top, [X|XS]);
				false -> happy(X + 1,Top, XS)@\label{lst:bug_2}@
			end
	end.
	
-spec main(pos_integer(),pos_integer()) -> 
	[pos_integer()].
main(N, M) -> 
	happy(N, M, []).
\end{lstlisting}
\end{multicols}
\end{minipage}

\begin{figure*}[h!]
\noindent\begin{minipage}{\linewidth}
\begin{lstlisting}[basicstyle=\ttfamily\scriptsize, frame=single, caption={SecEr reports that no discrepancies exist}, label=lst:correct,language=none,escapechar=@]
$ ./secer -f happy0.erl -li @\ref{lst:poi_happy0}@  -var Happy -oc 1 
          -f happy1.erl -li @\ref{lst:poi_happy1}@ -var Happy -oc 1 
          -funs [main/2] -to 15
          
Function: main/2
----------------------------
Generated test cases: 812
Both versions of the program generate identical traces for the defined points of interest
----------------------------
\end{lstlisting}
\end{minipage}
\end{figure*}

First of all, and before using SecEr, we have to slightly adapt these codes. 
Our implementation needs a common input function, but in the original {\tt happy0} module there is a function \texttt{main/0}, which just runs a concrete test. Therefore, we have replaced this function with \texttt{main/2} (see Listing \ref{lst:happy0}) to match it with the corresponding one in the {\tt happy1} module (Listing \ref{lst:happy1}). 
% In order to unify their interfaces, we have replaced \texttt{main/0} with \texttt{main/2} in the {\tt happy0} module (Listing \ref{lst:happy0}) making it applicable for a more general case. 
Moreover, in the {\tt happy1} module, we have added the \texttt{Happy} variable\footnote{This variable addition will not be needed when SecEr allows to define expressions as POIs.} (line \ref{lst:poi_happy1}), which stores the result of {\texttt{is_happy(X,[])}. In both modules, we have added a type specification (represented with {\tt spec} in Erlang) 
%in order to faster obtaining representative test cases
to obtain more useful test cases in less time\footnote{Note that depending on the ITC, the execution of these programs can loop forever. Although we run each execution of a ITC with a limited timeout, these useless ITCs are spending valuable time that could be used by more useful ITCs.}.
%in order to obtain more representative test cases\david{Aqui da la impresion de que se obtienen test cases peores sin el spec. Basta con decir que la tecnica se puede complementar con el spec para acelerar la busqueda de test cases de calidad}\tama{Es que es la verdad. se obitenen casos peores. No escondemos nada. La footnote lo intenta explicar.}\david{Solo digo de orientarlo hacia algo positivo en vez de hacia algo negativo} in less time.\footnote{Note that depending on the ITC, the execution of these programs can loop forever. Although we run each execution of a ITC with a limited timeout, these useless ITCs are spending valuable time that could be used by more useful ITCs.}

Listing \ref{lst:correct} shows the SecEr's output when comparing both implementations of the program using a timeout of 15 seconds. The POI for both modules is the {\tt Happy} variable, which is evaluated many times per execution, returning a trace with more than one element. As we can see, the execution of both implementations behave identically with respect to the \texttt{Happy} variable in the 812 generated test cases. 

In order to see the output of the tool when the behaviour of the two compared programs differ, we have introduced two different errors in the {\tt happy1} module of Listing \ref{lst:happy1}.
%\david{Aqui me ha parecido que ibamos a saltar al segundo caso de uso. En ningun momento se ha dicho que se iban a hacer experimentos con el codigo fuente de cada caso de uso, o si?}\tama{Arriba se introducen los casos de uso y se les asigna subseccions (ahora, cuando lo leiste no estaba). Respecto a experimentar se puede meter alguna frase general en la intro de la seccion si estamos todos de acuerdo}

\begin{description}

\begin{figure*}[p!]
\noindent\begin{minipage}{\linewidth}
\begin{lstlisting}[basicstyle=\ttfamily\scriptsize, frame=single, caption={SecEr reports discrepancies for the error inside function \texttt{is_happy} using variable \texttt{Happy} as POI.}, label=lst:errorinside1,language=none, escapechar=@]
Function: main/2
----------------------------
Generated test cases: 759
Mismatching test cases: 66 (8.69%)
    POIs comparison:
	+ {{'happy0.erl',@\ref{lst:poi_happy0}@,{var,'Happy'},1},
           {'happy1.erl',@\ref{lst:poi_happy1}@,{var,'Happy'},1}}
		 Unexpected trace value => 66 Errors
		 Example call: main(2,6)
------ Detected Error ------
Call: main(2,6)
Error Type: Unexpected trace value
POI: ({'happy0.erl',@\ref{lst:poi_happy0}@,{var,'Happy'},1}) trace:
[false,false,false,false,false,true,false,false,true,false,false,true,false,false,false,false,
false,true,false,false,false,true,false,false,false,false,true]
POI: ({'happy1.erl',@\ref{lst:poi_happy1}@,{var,'Happy'},1}) trace:
[false,false,false,false,false,false,false,false,true,false,false,true,false,false,false,false,
false,true,false,false,false,true,false,false,false,false,true,false,false,true]
----------------------------
\end{lstlisting}

\begin{lstlisting}[basicstyle=\ttfamily\scriptsize, frame=single, caption={SecEr reports discrepancies for the error inside function \texttt{is_happy} using variable \texttt{X} as POI.}, label=lst:errorinside2,language=none, escapechar=@]
Function: main/2
----------------------------
Generated test cases: 905
Mismatching test cases: 105 (11.6%)
    POIs comparison:
	+ {{'happy0.erl',@\ref{lst:poi_happy0}@,{var,'X'},1},
           {'happy1.erl',@\ref{lst:poi_happy1}@,{var,'X'},1}}
		 The second trace is longer => 105 Errors
		 Example call: main(6,3)
------ Detected Error ------
Call: main(6,3)
Error Type: The length of both traces differs
POI: ({'happy0.erl',@\ref{lst:poi_happy0}@,{var,'X'},1}) trace:
	 [6,7,8,9,10,11,12,13]
POI: ({'happy1.erl',@\ref{lst:poi_happy1}@,{var,'X'},1}) trace:
	 [6,7,8,9,10,11,12,13,14,15,16,17,18,19]
----------------------------
\end{lstlisting}
\end{minipage}
\end{figure*}

%\david{Cambiar "Detected error" en los listings por "Sample detected error"?}

\item[An error inside the \texttt{is_happy} function] The first error is introduced by replacing the whole line \ref{lst:bug_1} with {\small\texttt{X < 10 -> false;}}. % in the {\small\texttt{is_happy/2}} function. 
With this change, the behaviour of both programs differs. When the user runs SecEr using the previous POI, it produces the report shown in Listing \ref{lst:errorinside1}. SecEr is reporting that the POI has been executed several times and in some of these executions the values of the POI differed. Concretely, according to the given example, the executions of ITC \texttt{main(2,6)} compute different values at the sixth time both POIs are executed. Given this report, the user can continue testing using other POIs in order to find the source of the discrepancy. Since variable \texttt{Happy} is the result of calling the same function in both codes, there are two possible sources of the discrepancy: either the arguments that can take different values during the execution (i.e. \texttt{X}) or the code executed by the called function (i.e. \texttt{is_happy}). Following this idea, the user could use argument \texttt{X} as the new POI. 
In this case, if both versions produce different traces for \texttt{X}, then the discrepancy source must be in the parts of the code responsible for the values taken by variable \texttt{X}. 
Otherwise, the argument \texttt{X} can be discarded as the  source of the discrepancy, and the source must be in something executed by the \texttt{is_happy} function. 
Listing \ref{lst:errorinside2} shows the report provided by SecEr when selecting variable \texttt{X} as the POI. The reported discrepancy indicates that both traces are the same until a point in the execution were the version in Listing \ref{lst:happy1} continues producing values. This behaviour is the expected one because the result of \texttt{is_happy} has an influence in the number of times the call is executed. Therefore, the user can conclude that the arguments do not produce the discrepancy and the error is inside the \texttt{is_happy} function.

\item[An error outside the \texttt{is_happy} function]

\begin{figure*}[t!]
\noindent\begin{minipage}{\linewidth}
\begin{lstlisting}[basicstyle=\ttfamily\scriptsize, frame=single, caption={SecEr reports discrepancies for the error outside function \texttt{is_happy} using variable \texttt{Happy} as POI.}, label=lst:erroroutside1,language=none,escapechar=@]
Function: main/2
----------------------------
Generated test cases: 830
Mismatching test cases: 798 (96.14%)
    POIs comparison:
	+ {{'happy0.erl',@\ref{lst:poi_happy0}@,{var,'Happy'},1},
           {'happy1.erl',@\ref{lst:poi_happy1}@,{var,'Happy'},1}}
		 Unexpected trace value => 798 Errors
		 Example call: main(66,6)
------ Detected Error ------
Call: main(66,6)
Error Type: Unexpected trace value
POI: ({'happy0.erl',@\ref{lst:poi_happy0}@,{var,'Happy'},1}) trace:
[false,false,true,false,true,false,false,false,false,false,false,false,false,true,false,false,
true,false,false,false,true,false,false,false,false,true]
POI: ({'happy1.erl',@\ref{lst:poi_happy1}@,{var,'Happy'},1}) trace:
[false,true,false,false,false,false,false,true,false,true,false,false,false,false,true,false,
true,false,true]
----------------------------
\end{lstlisting}

\begin{lstlisting}[basicstyle=\ttfamily\scriptsize, frame=single, caption={SecEr reports discrepancies for the error outside function \texttt{is_happy} using variable \texttt{X} as POI.}, label=lst:erroroutside2,language=none,escapechar=@]
Function: main/2
----------------------------
Generated test cases: 787
Mismatching test cases: 766 (97.33%)
    POIs comparison:
	+ {{'happy0.erl',@\ref{lst:poi_happy0}@,{var,'X'},1},
           {'happy1.erl',@\ref{lst:poi_happy1}@,{var,'X'},1}}
		 Unexpected trace value => 766 Errors
		 Example call: main(63,3)
------ Detected Error ------
Call: main(63,3)
Error Type: Unexpected trace value
POI: ({'happy0.erl',@\ref{lst:poi_happy0}@,{var,'X'},1}) trace:
	 [63,64,65,66,67,68,69,70,71,72,73,74,75,76,77,78,79]
POI: ({'happy1.erl',@\ref{lst:poi_happy1}@,{var,'X'},1}) trace:
	 [63,65,67,69,71,73,75,77,79,80,82,83,85,87,89,91]
----------------------------
\end{lstlisting}
\end{minipage}
\end{figure*}

The second error consists in the replacement of the whole line \ref{lst:bug_2} with {\small\texttt{false -> happy(X + 2, Top, XS)}}. % in the {\small\texttt{happy/3}} function. %and recursively executing the point of interest, .
The output of SecEr in this case is depicted in Listing \ref{lst:erroroutside1}. The discrepancy reported is exactly the same one than the one reported in the previous error. Therefore, the user could use here variable \texttt{X} as the next POI following the same idea explained above. Listing \ref{lst:erroroutside2} shows the report of SecEr in this case. Here the error is different from the one obtained above. It indicates that the values computed for \texttt{X} differ between versions. This is a clear indicator that this error is not being produced by function \texttt{is_happy} and it comes from where the variable \texttt{X} obtains its values. There are only two places in the code where this occurs (apart from the initial call that according to the reported error seems to be correctly defined), so now it is easy for the user to spot the error using this information. 
%\sergio{Hay que tener cuidado aqui por lo de los tipos de errores. Nose si vale la pena ponerlos todos. En estos casos descritos el error siempre es del tipo unexpected trace value, tenga o no la misma longitud la traza, porque es el primero que se detecta. Si queremos mencionar la longitud de las trazas yo utilizaria otro ejemplo (string) donde aparecen trazas vacias y tal.}

\end{description}

\setlength{\columnsep}{1cm}
\begin{figure}[b!]
Original version
\hrule
\begin{multicols}{2}
\begin{lstlisting}[tabsize=2,basicstyle=\ttfamily\tiny, 
%frame=bt, 
%caption={string.erl (old version)}, label=lst:string0, 
numbers=left, stepnumber=1,escapechar=@]
tokens(S, Seps) ->
    Res = tokens1(S, Seps, []).@\label{lst:string0_poi}@
tokens1([C|S], Seps, Toks) ->
  case member(C, Seps) of
  	true -> 
      tokens1(S, Seps, Toks);
  	false -> 
      tokens2(S, Seps, Toks, [C])
  end;
tokens1([], _Seps, Toks) ->
  reverse(Toks).@\columnbreak@
tokens2([C|S], Seps, Toks, Cs) ->
  case member(C, Seps) of
  	true -> 
      tokens1(S, Seps, [reverse(Cs)|Toks]);
  	false -> 
      tokens2(S, Seps, Toks, [C|Cs])
  end;
tokens2([], _Seps, Toks, Cs) ->
    reverse([reverse(Cs)|Toks]).
\end{lstlisting}
\end{multicols}
%\hrule
Optimized version
\hrule
\begin{multicols}{2}
\begin{lstlisting}[tabsize=2,basicstyle=\ttfamily\tiny, 
%frame=bt,
%caption={string.erl\\ (optimized version)}, label=lst:string1, 
numbers=left, stepnumber=1,escapechar=@]
tokens(S, Seps) ->
  Res =  @\label{lst:string1_poi1}@
    case Seps of
      [] ->@\label{lst:string1_empty}@
        case S of
          [] -> [];
          [_|_] -> [S]
        end;
      [C] ->
        Res1 = tokens_single_1(reverse(S), C, []);@\label{lst:string1_poi2}@
      [_|_] ->
        Res2 = tokens_multiple_1(reverse(S), Seps, [])@\label{lst:string1_poi3}@
    end.
tokens_single_1([Sep|S], Sep, Toks) ->
  tokens_single_1(S, Sep, Toks);
tokens_single_1([C|S], Sep, Toks) ->
  tokens_single_2(S, Sep, Toks, [C]);
tokens_single_1([], _, Toks) ->
  Toks.
tokens_single_2([Sep|S], Sep, Toks, Tok) ->
  tokens_single_1(S, Sep, [Tok|Toks]);
tokens_single_2([C|S], Sep, Toks, Tok) ->
  tokens_single_2(S, Sep, Toks, [C|Tok]);
tokens_single_2([], _Sep, Toks, Tok) ->
  [Tok|Toks].@\columnbreak@
tokens_multiple_1([C|S], Seps, Toks) ->
  case member(C, Seps) of
    true -> 
      tokens_multiple_1(S, Seps, Toks);
    false -> 
      tokens_multiple_2(S, Seps, Toks, [C])@\label{lst:string1_error}@
  end;
tokens_multiple_1([], _Seps, Toks) ->
  Toks.
tokens_multiple_2([C|S], Seps, Toks, Tok) ->
  case member(C, Seps) of
    true -> 
      tokens_multiple_1(S, Seps, [Tok|Toks]);
    false -> 
      tokens_multiple_2(S, Seps, Toks, [C|Tok])
  end;
tokens_multiple_2([], _Seps, Toks, Tok) ->
  [Tok|Toks].
\end{lstlisting}
\end{multicols}
\hrule
\caption{string.erl (original and optimized versions)}
\label{fig:string1}
\end{figure}

As we can observe in both errors, it does not matter where the bug is located as long as the bug affects the values computed at the POI.  As we have seen in the examples, another interesting feature of the tool arises when we have a POI that is evaluated several times during the execution of the program. In this case, SecEr allows us to differentiate between two kinds of errors: traces that differ in their number of elements where a trace is a prefix of the other one (Listing \ref{lst:errorinside2}) and traces with the same values in different order or with different values (Listing \ref{lst:erroroutside2}).
%\textbf{El primer tipo de error fa que simplement hi hasca una discrepancia en el tamany de valors trasats per X (el error esta dins de happy perque els valors per a la mateixa X diferisen) , mentre que el segon fa que els valors cambien (el error be de fora de is_happy) }

\section{Use case 2: An improvement of the \texttt{string:tokens/2} function}\label{sec:uc_string}

In this case of study, we consider a real commit of the Erlang/OTP distribution that improved the performance of the \texttt{string:tokens/2} function. Figure \ref{fig:string1} shows the code of the original and the improved versions. The differences introduced in this commit can be consulted here:

\begin{center}
\scriptsize
\url{https://github.com/erlang/otp/commit/53288b441ec721ce3bbdcc4ad65b75e11acc5e1b}
\end{center}

The improvement performed in this commit consists in two main changes. The first one is a general improvement obtained by reversing the input string (the one that is going to be tokenized) at the beginning of the process. The second one improves the cases where the separator list has only one element. The algorithm uses two auxiliary functions in both cases, so its structure is kept between both versions. However, the optimized code defines two versions of these functions to cover the single-element list of separators and the rest of cases separately.

We can use SecEr to check whether the behaviour of both versions is the same. A good way to start is to check whether the results produced by function \texttt{tokens/2} are the same. In order to do this, we have to slightly modify the codes by adding variable \texttt{Res} in line \ref{lst:string0_poi} of the original version and in line \ref{lst:string1_poi1} of the improved version\footnote{Note that this modification is only needed because of the current limitations of the tool. However, the approach has not this limitation, as it allows to select any expression, not only variables.}, both in Figure \ref{fig:string1}. Then, this variable is used as a POI. This allow the user to check that both versions preserve the same behaviour (see Listing~\ref{lst:stringcorrect}). 

%\tama{aqui run de secer con un solo POI cuando esta todo bien.}

\begin{figure*}[th!]
\noindent\begin{minipage}{\linewidth}
\begin{lstlisting}[basicstyle=\ttfamily\scriptsize, frame=single, caption={SecEr reports that no discrepancies exist}, label=lst:stringcorrect,language=none, escapechar=@]
$ ./secer -f string0.erl -li @\ref{lst:string0_poi}@ -var Res -oc 1 
          -f string1.erl -li @\ref{lst:string1_poi1}@ -var Res -oc 1 
          -funs [main/2] -to 15
          
Function: tokens/2
----------------------------
Generated test cases: 7044
Both versions of the program generate identical traces for the defined points of interest
----------------------------
\end{lstlisting}
\end{minipage}
%\end{figure*}
%
%\begin{figure*}[t!]
\noindent\begin{minipage}{\linewidth}
\begin{lstlisting}[basicstyle=\ttfamily\scriptsize, frame=single, caption={SecEr reports discrepancies after modifying optimized string.erl}, label=lst:stringError1, language=none, escapechar=@]
Function: tokens/2
----------------------------
Generated test cases: 6576
Mismatching test cases: 5040 (76.64%)
    POIs comparison:
	+ {{'string0.erl',@\ref{lst:string0_poi}@,{var,'Res'},1},
           {'string1.erl',@\ref{lst:string1_poi1}@,{var,'Res'},1}}
		 Unexpected trace value => 5040 Errors
		 Example call: tokens([12,4,5],[2,3,2,5,0,1])
------ Detected Error ------
Call: tokens([12,4,5],[2,3,2,5,0,1])
Error Type: Unexpected trace value
POI: ({'string0.erl',@\ref{lst:string0_poi}@,{var,'Res'},1}) trace:
	 [[[12,4]]]
POI: ({'string1.erl',@\ref{lst:string1_poi1}@,{var,'Res'},1}) trace:
	 [[4,[12,4]]]
---------------------
\end{lstlisting}
\end{minipage}
\end{figure*}

%\begin{figure*}[t!]
%\noindent\begin{minipage}{\linewidth}
%\begin{lstlisting}[basicstyle=\ttfamily\scriptsize, frame=single, caption={SecEr reports that no discrepancies exist}, label=lst:stringcorrect,language=none]
%$ ./secer ...
%          
%Function: tokens/2
%----------------------------
%Generated test cases: 12861
%Both versions of the program generate identical traces for the defined 
%points of interest
%----------------------------
%\end{lstlisting}
%\end{minipage}
%\end{figure*}

We can now consider a hypothetical scenario where an error was introduced in the aforementioned commit. Suppose that line \ref{lst:string1_error} in Figure \ref{fig:string1} (optimized version) is replaced by the following expression:

\begin{center}
\scriptsize
\texttt{[C | tokens_multiple_2(S, Seps, Toks, [C])]}
\end{center}

\begin{figure*}[th!]
\noindent\begin{minipage}{\linewidth}
\begin{lstlisting}[basicstyle=\ttfamily\scriptsize, frame=single, caption={SecEr report using variable \texttt{Res1} as the POI for the optimized version.}, label=lst:stringError1_1,language=none, escapechar=@]
Function: tokens/2
----------------------------
Generated test cases: 6570
Mismatching test cases: 6132 (93.33%)
    POIs comparison:
	+ {{'string0.erl',@\ref{lst:string0_poi}@,{var,'Res'},1},
           {'string1.erl',@\ref{lst:string1_poi2}@,{var,'Res1'},1}}
		 The second trace is empty => 6132 Errors
		 Example call: tokens([0,5,2],[])
------ Detected Error ------
Call: tokens([0,5,2],[])
Error Type: The second trace is empty
POI: ({'string0.erk',@\ref{lst:string0_poi}@,{var,'Res'},1}) trace:
	 [[[0,5,2]]]
POI: ({'string1.erl',@\ref{lst:string1_poi2}@,{var,'Res1'},1}) trace:
	 []
----------------------------
\end{lstlisting}
\end{minipage}
%\end{figure*}
%
%\begin{figure*}[t!]
\noindent\begin{minipage}{\linewidth}
\begin{lstlisting}[basicstyle=\ttfamily\scriptsize, frame=single, caption={SecEr report using variable \texttt{Res2} as the POI for the optimized version.}, label=lst:stringError1_2, language=none, escapechar=@]
Function: tokens/2
----------------------------
Generated test cases: 7134
Mismatching test cases: 6649 (93.2%)
    POIs comparison:
	+ {{'string0.erl',@\ref{lst:string0_poi}@,{var,'Res'},1},
           {'string1.erl',@\ref{lst:string1_poi3}@,{var,'Res2'},1}}
		 The second trace is empty => 1348 Errors
		 Example call: tokens([2,5,6,6,0],[])
	+ {{'string0.erl',@\ref{lst:string0_poi}@,{var,'Res'},1},
           {'string1.erl',@\ref{lst:string1_poi3}@,{var,'Res2'},1}}
		 Unexpected trace value => 5301 Errors
		 Example call: tokens([9,10,2],[61,4,2,12,0,21,4,31])
------ Detected Error ------
Call: tokens([2,5,6,6,0],[])
Error Type: The second trace is empty
POI: ({'string0.erl',@\ref{lst:string0_poi}@,{var,'Res'},1}) trace:
	 [[[2,5,6,6,0]]]
POI: ({'string1.erl',@\ref{lst:string1_poi3}@,{var,'Res2'},1}) trace:
	 []
----------------------------
------ Detected Error ------
Call: tokens([9,10,2],[61,4,2,12,0,21,4,31])
Error Type: Unexpected trace value
POI: ({'string_old.erl',@\ref{lst:string0_poi}@,{var,'Res'},1}) trace:
	 [[[9,10]]]
POI: ({'string_error1.erl',@\ref{lst:string1_poi3}@,{var,'Res2'},1}) trace:
	 [[10,[9,10]]]
----------------------------
\end{lstlisting}
\end{minipage}
\end{figure*}

In this scenario, SecEr reports that some of the traces differ (see Listing~\ref{lst:stringError1}). The error indicates that the produced values are not the expected ones. As in this particular case the main modification is the specialization of function \texttt{tokens1/3}, analyze the differences between the new versions of this function seems a good way to continue searching discrepancies, i.e. \texttt{tokens_single_1/3} and \texttt{tokens_multiple_1/3}. For doing this, we need to add new variables \texttt{Res1} and \texttt{Res2} in lines \ref{lst:string1_poi2} and \ref{lst:string1_poi3} of Figure \ref{fig:string1} (optimized version). The report for both POIs is depicted in Listings \ref{lst:stringError1_1} and \ref{lst:stringError1_2}, respectively. Let us first analyze the report in Listing \ref{lst:stringError1_1}. At first sight, one can think that we have spotted the function responsible of the discrepancy, but this is not true. What SecEr is reporting here is that for some cases, the optimized version has not produced any value. For instance, the call \texttt{tokens([0,5,2],[])}, that SecEr uses to exemplify the problem,  is executed by line \ref{lst:string1_empty} of the optimized version, so it is not entering to the case branch that executes \texttt{tokens_single_1/3}. Similar errors arise when the separator list contains more than one element, executing in this case the branch in line \ref{lst:string1_poi3}. So this kind of errors are expected and are not considered to be a discrepancy. On the other hand, the report on Listing \ref{lst:stringError1_2} shows two types of discrepancies. The first one has the same explanation of the one reported in Listing \ref{lst:stringError1_1}. However, the second one corresponds to a real source of discrepancy, since they are cases where \texttt{tokens_multiple_1/3} is executed and produced different results than \texttt{tokens1/3}. Note that almost 75\% (5301 out of 7134) of the ITCs produced by SecEr correspond to the relevant error. This is due to the priority given to the discrepant traces as it is  explained in Section 3.3 of \cite{insa2017LOPSTR_LNCS}. Listing \ref{lst:stringError1_2} also reveals one interesting feature of SecEr, which is that it can report different types of discrepancies for a given pair of POIs and provide the user with an example of each of them.

\section{Use case 3: Complex numbers}\label{sec:uc_complex}

\setlength{\columnsep}{1cm}
\begin{figure}%[t!]
%\hrule
%Optimized version
\hrule
\begin{multicols}{2}
\begin{lstlisting}[tabsize=2,basicstyle=\ttfamily\tiny, 
%frame=bt,
%caption={string.erl\\ (optimized version)}, label=lst:string1, 
numbers=left, stepnumber=1,escapechar=@]
-record(complex, {real, img}).

calculate(AR, AI, BR, BI) ->
    A = #complex{real=AR, img=AI},
    B = #complex{real=BR, img=BI},
    Sum = add (A, B),
    Product = multiply (A, B),
    Negation = negation (A),
    Inversion = inverse (A),@\label{lst:complex_poi2}@
    Conjugate = conjugate (A),
    {Sum, Product, Negation, Inversion, Conjugate}.@\label{lst:complex_poi1}@
 
add (A, B) ->
    RealPart = A#complex.real + B#complex.real,
    ImgPart = A#complex.img + B#complex.img,
    #complex{real=RealPart, img=ImgPart}.
 
multiply (A, B) ->
    RealPart = (A#complex.real * B#complex.real) 
        - (A#complex.img * B#complex.img),
    ImgPart = (A#complex.real * B#complex.img) 
        + (B#complex.real * A#complex.img),
    #complex{real=RealPart, img=ImgPart}.
 
negation (A) ->
    #complex{real=-A#complex.real, img=-A#complex.img}.
 
inverse (A) ->
    C = conjugate (A),
    % @\color{red}{\emph{Mod = (A\#complex.real * A\#complex.real)}@ @\label{lst:complex_right1}@
    % @\color{red}{\emph{   + (A\#complex.img * A\#complex.img),}@ %RIGHT @\label{lst:complex_right2}@
    Mod = (A#complex.real * A#complex.img) @\label{lst:complex_bug1}@
        + (A#complex.img * A#complex.img), %WRONG @\label{lst:complex_bug2}@
    RealPart = C#complex.real / Mod,@\label{lst:complex_poi4}@
    ImgPart = C#complex.img / Mod,
    #complex{real=RealPart, img=ImgPart}.@\label{lst:complex_poi3}@
 
conjugate (A) ->
    RealPart = A#complex.real,
    ImgPart = -A#complex.img,
    #complex{real=RealPart, img=ImgPart}.
\end{lstlisting}
\end{multicols}
\vspace{0.25cm}
\hrule
\caption{complex1.erl}
\label{fig:complex}
\end{figure}

The following use case is extracted\footnote{It is, in turn, built using a module from the Rosetta Code repository:\\ \url{http://rosettacode.org/wiki/Arithmetic/Complex}} from the use-case database of \texttt{EDD}\footnote{\url{https://github.com/tamarit/edd}}, an algorithmic debugger for Erlang:

\begin{center}
\footnotesize
\url{https://github.com/tamarit/edd/tree/master/examples/complex1}.
\end{center}

\begin{figure*}[ht!]
\noindent\begin{minipage}{\linewidth}
\begin{lstlisting}[basicstyle=\ttfamily\scriptsize, frame=single, caption={Both versions behave in the same way for variable \texttt{Negation}}, label=lst:complex_secer_1,language=none,escapechar=@]
$ ./secer -f complex0.erl -li @\ref{lst:complex_poi1}@  -var Negation -oc 1 
          -f complex1.erl -li @\ref{lst:complex_poi1}@ -var Negation -oc 1 
          -funs [main/2] -to 15
 
Function: calculate/4
----------------------------
Generated test cases: 6459
Both versions of the program generate identical traces for the defined points of interest
----------------------------
\end{lstlisting}
\end{minipage}
%\end{figure*}
%
%
%\begin{figure*}[t!]
\noindent\begin{minipage}{\linewidth}
\begin{lstlisting}[basicstyle=\ttfamily\scriptsize, frame=single, caption={Discrepancies reported for variable \texttt{Inversion}}, label=lst:complex_secer_2,language=none,escapechar=@]
Function: calculate/4
----------------------------
Generated test cases: 6963
Mismatching test cases: 6844 (98.29%)
    POIs comparison:
	+ {{'complex0.erl',@\ref{lst:complex_poi1}@,{var, 'Inversion'},1},
           {'complex1.erl',@\ref{lst:complex_poi1}@,{var, 'Inversion'},1}}
		 The second trace is empty => 170 Errors
		 Example call: calculate(-75,0,38,6)
	+ {{'complex0.erl',@\ref{lst:complex_poi1}@,{var, 'Inversion'},1},
           {'complex1.erl',@\ref{lst:complex_poi1}@,{var, 'Inversion'},1}}
		 Unexpected trace value => 6674 Errors
		 Example call: calculate(-23,11.147602744042517,-13.290018732881684,
		 		 7.818912260066709)
------ Detected Error ------
Call: calculate(-75,0,38,6)
Error Type: The first trace is empty
POI: ({'complex1.erl',@\ref{lst:complex_poi1}@,{var, 'Inversion'},1}) trace:
	 [{complex,-0.013333333333333334,0.0}]
POI: ({'complex0.erl',@\ref{lst:complex_poi1}@,{var, 'Inversion'},1}) trace:
	 []
----------------------------
------ Detected Error ------
Call: calculate(-23,11.147602744042517,-13.290018732881684,7.818912260066709)
Error Type: Unexpected trace value
POI: ({'complex0.erl',@\ref{lst:complex_poi1}@,{var, 'Inversion'},1}) trace:
	 [{complex,-0.03520754596864929,-0.017064336350048604}]
POI: ({'complex1.erl',@\ref{lst:complex_poi1}@,{var, 'Inversion'},1}) trace:
	 [{complex,0.1740765027306382,0.08437111737014728}]
----------------------------
\end{lstlisting}
\end{minipage}
\end{figure*}

\begin{figure*}[p!]
\noindent\begin{minipage}{\linewidth}
\begin{lstlisting}[basicstyle=\ttfamily\scriptsize, frame=single, caption={Discrepancies reported for variable \texttt{RealPart}}, label=lst:complex_secer_3,language=none,escapechar=@]
Function: calculate/4
----------------------------
Generated test cases: 7326
Mismatching test cases: 7192 (98.17%)
    POIs comparison:
	+ {{'complex0.erl',@\ref{lst:complex_poi3}@,{var,'RealPart'},1},
           {'complex1.erl',@\ref{lst:complex_poi3}@,{var,'RealPart'},1}}
		 The second trace is empty => 188 Errors
		 Example call: calculate(11,0,-2.769732522433009,-1.9949121299611243)
	+ {{'complex0.erl',@\ref{lst:complex_poi3}@,{var,'RealPart'},1},
           {'complex1.erl',@\ref{lst:complex_poi3}@,{var,'RealPart'},1}}
		 Unexpected trace value => 7004 Errors
		 Example call: calculate(-206.34436567989434,15.930666184706574,1.6000469126136903,
		 		 6.147738016916419)
------ Detected Error ------
Call: calculate(11,0,-2.769732522433009,-1.9949121299611243)
Error Type: The second trace is empty
POI: ({'complex0.erl',@\ref{lst:complex_poi3}@,{var,'RealPart'},1}) trace:
	 [0.09090909090909091]
POI: ({'complex1.erl',@\ref{lst:complex_poi3}@,{var,'RealPart'},1}) trace:
	 []
----------------------------
------ Detected Error ------
Call: calculate(-206.34436567989434,15.930666184706574,1.6000469126136903,6.147738016916419)
Error Type: Unexpected trace value
POI: ({'complex0.erl',@\ref{lst:complex_poi3}@,{var,'RealPart'},1}) trace:
	 [-0.004817552514292979]
POI: ({'complex1.erl',@\ref{lst:complex_poi3}@,{var,'RealPart'},1}) trace:
	 [0.06802373692423881]
----------------------------
\end{lstlisting}
\end{minipage}
%\end{figure*}
%
%\begin{figure*}%[b!]
\noindent\begin{minipage}{\linewidth}
\begin{lstlisting}[basicstyle=\ttfamily\scriptsize, frame=single, caption={No discrepancies reported for variable \texttt{C}}, label=lst:complex_secer_4,language=none]
Function: calculate/4
----------------------------
Generated test cases: 7272
Both versions of the program generate identical traces for the defined points of interest
----------------------------
\end{lstlisting}
%\end{minipage}
%\end{figure*}
%
%
%\begin{figure*}[h!]
%\noindent\begin{minipage}{\linewidth}
\begin{lstlisting}[basicstyle=\ttfamily\scriptsize, frame=single, caption={Discrepancies reported for variable \texttt{Mod}}, label=lst:complex_secer_5,language=none, escapechar=@]
Function: calculate/4
----------------------------
Generated test cases: 7489
Mismatching test cases: 7352 (98.17%)
    POIs comparison:
	+ {{'complex0.erl',@\ref{lst:complex_poi4}@,{var,'Mod'},1},
           {'complex1.erl',@\ref{lst:complex_poi4}@,{var,'Mod'},1}} 
		 Unexpected trace value => 6706 Errors
		 Example call: calculate(24,-10.509620259997908,14,-18)
------ Detected Error ------
Call: calculate(24,-10.509620259997908,14,-18)
Error Type: Unexpected trace value
POI: ({'complex0.erl',@\ref{lst:complex_poi4}@,{var,'Mod'},1}) trace:
	 [686.4521180093585]
POI: ({'complex1.erl',@\ref{lst:complex_poi4}@,{var,'Mod'},1}) trace:
	 [-141.77876823059128]
----------------------------
\end{lstlisting}
\end{minipage}
\end{figure*}

This use case is formed by two versions of the same module that are almost identical except that one of them has a bug. Figure \ref{fig:complex} shows the code (as it is on GitHub) of the buggy version. The bug is in lines \ref{lst:complex_bug1} and \ref{lst:complex_bug2}, while the correct code is obtained by replacing these lines with lines \ref{lst:complex_right1} and \ref{lst:complex_right2}. In the rest of the section, we show how SecEr can be used to detect this type of bugs.

We can start our test by choosing as a POI each variable of the resulting tuple of function \texttt{calculate/4} in line \ref{lst:complex_poi1}, one by one. We named \texttt{complex0} and \texttt{complex1}, the correct and the buggy versions of the code, respectively. The report given by SecEr for all the variables is similar to the one depicted in Listing \ref{lst:complex_secer_1}, which shows that both versions behave in the same way, except for variable \texttt{Inversion} for which SecEr indicates that some discrepancies exist as it is shown in Listing \ref{lst:complex_secer_2}. In fact, this listing is showing two kinds of errors. The first indicates the lack of traces for one of the executions  (probably due to runtime errors) and the second references to the discrepancy of the computed values. 
%This first error indicates that some executions have raised runtime errors. 
%In this case, we know that they are due to a division by zero in line \ref{lst:complex_poi4}, but the user does not know anything yet about the source of this discrepancy\sergio{Esta frase a lo mejor la borraria. Es como un spoiler de la ostia, seria mejor dejar al usuario ver como poco a poco nos acercamos al error y por que salia el error de antes. Diria que nos vamos a centrar en el otro porque al ser traza vacia sera algo que falla antes de llegar al POI o algo asi para justificarlo}.\tama{Entiendo lo que dices pero queda raro dejar esto abierto y comentarlo y (lo que mas me preocupa) referenciarlo mas tarde} 
%Therefore, the user should continue searching for it. 
Since variable \texttt{Inversion} comes from the result of the call to function \texttt{inverse/1} in line \ref{lst:complex_poi2}, there are two possible sources of the discrepancy: the argument (variable \texttt{A} in this case) or the definition of function \texttt{inverse/1}. Therefore, the second step could be to check whether both versions are using the same arguments by placing a POI in variable \texttt{A}. However, this variable is used in all the other calls of function \texttt{calculate/4}, so this means that \texttt{A} most likely is not the source of the discrepancy. Therefore, the user could decide to continue searching for the bug inside function \texttt{inverse/1} using as a POI the variables that defines the result, i.e. \texttt{RealPart} and \texttt{ImgPart}, both in line \ref{lst:complex_poi3}. Listing \ref{lst:complex_secer_3} shows the report provided by SecEr for variable \texttt{RealPart}. At this point, it is not needed to rerun SecEr using as a POI variable \texttt{ImgPart}, because we have already found an error. In this concrete case, if SecEr is rerun, a similar report will be produced. So, the next step is to place a POI in one of the two variables that define the value of \texttt{RealPart}, i.e. variables \texttt{C} and \texttt{Mod} in line \ref{lst:complex_poi4}. The report for variable \texttt{C} is depicted in Listing \ref{lst:complex_secer_4}, whereas for variable \texttt{Mod} is in Listing \ref{lst:complex_secer_5}. Thanks to these last reports, the user already can easily find the source of the discrepancy in the assignment of variable \texttt{Mod}. 
%\sergio{Los listings \ref{lst:complex_secer_2} y \ref{lst:complex_secer_3} tiene una cantidad incorrecta de errores en el mensaje de informacion. Con la ultima version de secer ya se separan los errores correctamente. Deben salir 2 tipos de error. Uno es con la traza vacia y otro con unexpected trace value} 

% Bibliografia
\bibliography{\biblio{biblio}}
\bibliographystyle{abbrv}

%% Termina el documento
\end{document}